\documentstyle[11pt,epsfig]{article}
\newcommand{\beq}{\begin{equation}}
\newcommand{\eeq}{\end{equation}}

\columnwidth\textwidth
\begin{document}
\begin{flushright}
BUTP-96 \\
%hep-ph/96xxxxx
\end{flushright}

\begin{center}
\Large{ \bf A Quasi-maximal mixing ansatz for neutrino oscillations.}\\
\end{center}

\begin{center}
E. Torrente-Lujan.\\
Inst. fur Theoretische Physik, Universitat Bern \\
Sidlerstrasse 5, 3012 Bern, Switzerland. \\
e-mail: e.torrente@cern.ch \\
\end{center}

\begin{abstract}
Inspired by the atmospheric multi-GeV neutrino data, we consider neutrino
flavor mixing matrices which are maximal 
in a $2\times 2$ subsector.
This condition is a strong restriction: the
full matrix including complex phases depends essentially 
on one parameter. The survival probability
is an universal function of $L/E$, independent of generation, in the region of
interest for the accelerator and atmospheric experiments. It is possible with the solar neutrino data alone to recover completely the mixing
matrix suggested by the atmospheric data.
 The results are not essentially modified by the MSW
effect in the solar data. Consequences for future experiments are considered.

%PACS: 

\end{abstract}

%\begin{center}
%{\em Submitted to Phys. Lett. B}
%\end{center}

\newpage

The hypothesis that the mixing in the Lepton sector is threefold maximal has
been discussed recently \cite{per1}. Such mixing would imply specific forms
for the Lepton mass matrices, corresponding to a cyclic permutation symmetry
among generations. The mixing is maximal in the sense that all the elements of the
mixing matrix have equal modulus. The intrinsic CP violation is also maximal. As a
consequence of this hypothesis the survival probabilities, as measured in
dissapareance experiments, $P_{ll}=P(\nu_l\to\nu_l)$ are identical. $P=P_{ll}$
is an universal  function of L/E, L the distance of propagation, E the neutrino
energy. In this model detailed  predictions depend on the character of the
neutrino squared  mass difference spectrum, a pronounced hierarchy must be
 assumed in order to accommodate the experimental data, particularly the solar data.

In \cite{bil1} and \cite{per2} it is shown that matter effects do not
contribute to the averaged values of the probabilities of transition of solar
neutrinos into other states in case of maximal mixing of any number of species.

Completely "democratic" mass matrices has been proposed already in the context of
the Seesaw mechanism (see \cite{koi1} and references therein). As it can be
easily shown maximal mixing appears as a possible particular case in
diagonalizing democratic mass matrices.

There are by now only two experimental groups which claim to have positive
evidence of neutrino oscillations, KAMIOKANDE \cite{fuku1} with its multi-GeV data
set and the LSND experiment \cite{atha1}.

A new analysis of the published  KAMIOKANDE data of multi-GeV
atmospheric neutrinos is presented in terms of three flavors
in \cite{yas1} .
The optimum set of parameters is found to be
 $(\Delta m^2_{21},\Delta m_{31}^2)$
$=(3.8\times 10^{-2}, 1.4\times 10^{-2})\ eV^2$, and
$(\theta_{12},\theta_{13},\theta_{23})$ $=(19^o, 43^o, 41^o)$.
$\theta_{ij}$ are the mixing angles in the standard 3 flavor parameterization
\cite{pdg1}.
The explicit mixing matrix corresponding to these angles is:
\begin{eqnarray}
u_{multi-GeV}=\pmatrix{0.69 & 0.23 & 0.68 \cr -0.67 & 0.57 & 0.48 \cr -0.27 &
-0.79 & 0.55 }
\label{e2020}
\end{eqnarray}

In \cite{yas1}, the minimization is
carried out over a sparse grid of the parameter space by CPU needs. Possible
complex phases has been neglected. 
A maximal mixing model with the standard mass hierarchy is hardly compatible
with this result even in a broad sense.

The LSND experiment observes $\overline{\nu}_e$ in excess from a 
proton beam dump at LAMPF. 
It corresponds to a oscillation  probability of
$(3.1\pm 1.0\pm 0.5)\times 10^{-3}$ \cite{atha1},
if the observed excess is interpreted as a 
$\overline{\nu}_\mu\to\overline{\nu}_e$ transition.

\begin{figure}
\centering\hspace{2cm}
{\epsfig{file=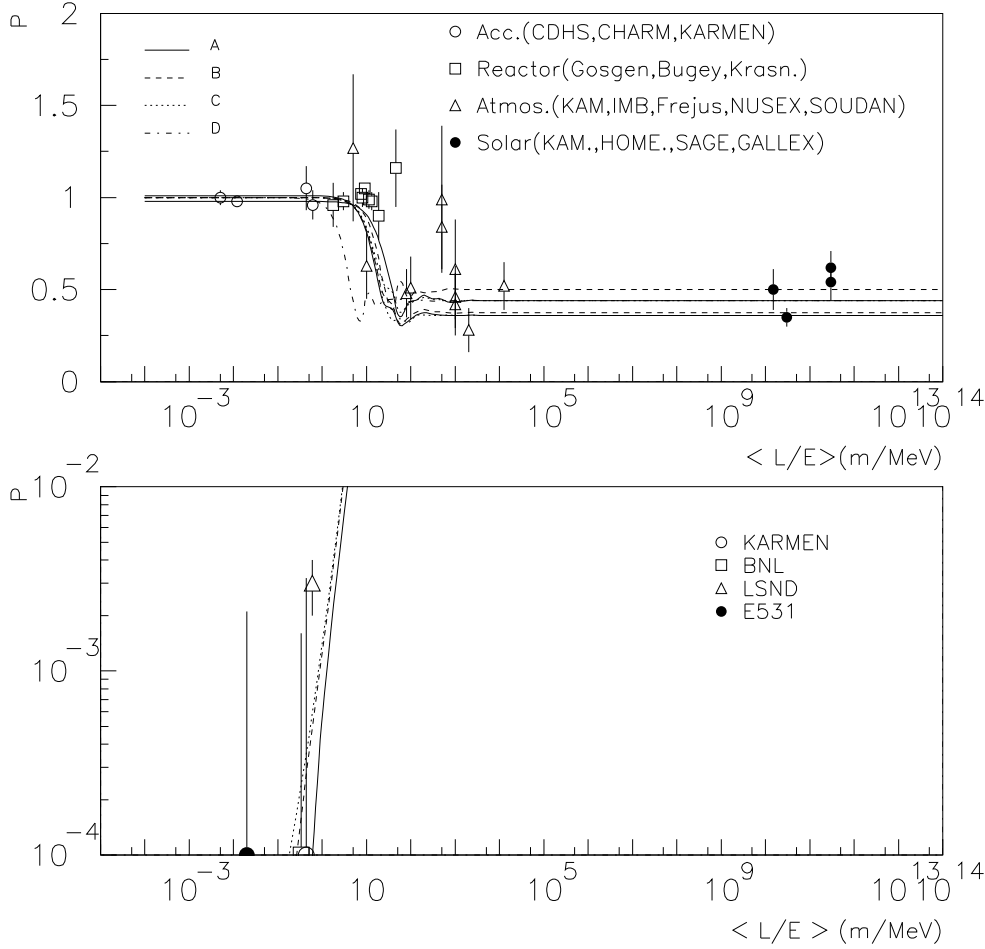,height=14cm}}
\caption{Top Figure. Individual data points: $P_{ll}$ measured in
diverse experiments. See Table (1) in
\protect\cite{per1} for detailed information about very individual point. The
atmospheric data points are shown with no corrections. 
Continuos and dashed lines: $P_{ee}^\infty,P_{\mu\mu}^\infty$ for different mixing matrices, the mass
differences are the ones obtained in \protect\cite{yas1} except for (D). (A) With the mixing
matrix (\protect\ref{e2020}).(B) With the matrix
 (\protect\ref{e1001}). (C) With the matrix (\protect\ref{e4001}). (D)
With the matrix (\protect\ref{e4001}) and 
 $(\Delta m^2_{21},\Delta m_{31}^2)$
$=(1.38\times 10^{-1}, 1.14\times 10^{-1})\ eV^2$.
Bottom Figure. Transition probabilities coming from different experiments (see
Table (\protect\ref{t1})). The different lines correspond to the case (C) referred
before except the dot-dashed one which correspond to case (D). The cases (A),(B) have been omitted for clarity. Their behavior is
essentially the same as case (C). }
\label{f1}
\end{figure}

In Fig.(\ref{f1}), we summarize virtually all the experimental limits avalaible
until now about  neutrino oscillations. A main observation can be drawn. There
are fundamentally only two differentiated regions in the $L/E$ line. 
This implies there is only one relevant mass difference scale 
($\Delta m^2_0\approx
10^{-1}-10^{-2} eV^2$). There are only two possibilities then: a)  
two of the generations are degenerated or nearly degenerated 
($\Delta m^2_1<10^{-11} eV^2$); b) the mass differences between all the
generations are approximately equal ($\approx \Delta m^2_0$) .

 For low L/E ($L/E<<1/\Delta m^2_0$)
, the survival probabilities are universal $P_{ee}\simeq
P_{\mu\mu}$. At high $L/E$, $P_{ee}$ is practically constant
within experimental errors. This is specially true if one discard 
the Homestake experiment (or atribute a bigger 
error to it).

The  interpretation of the 
 atmospheric data presents special problems. In one side, it is not possible to consider only
purely survival probabilities in 
general for three neutrino species. 
In another side, the
spread  in L and E are significantly larger and less well-known than in the rest
of experiments. 
A detailed analysis for each experiment is needed, out of the
scope of this work. 

Much less data is avalaible for transition probabilities, which are shown in
Fig.(\ref{f1}) ( bottom).

The object of this work is to show how we can recover a mixing matrix equivalent
to (\ref{e2020}), including in addition complex phases,
 from the information contained in
Figure (\ref{f1}) together with a highly predictive mixing matrix ansatz.

{\bf We are interested} in the case where the mixing is maximal in the $\mu-\tau$
sector, in the sense that all the matrix entries 
$u_{ij}, (2\leq i,j\leq 3)$, are
of equal magnitude. The cases where the mixing is maximal
in the $e-\mu$ or $e-\tau$ sectors are completely analogous and  will be treated later. Explicitly, we consider the matrix ansatz
\begin{eqnarray}
u&=&\frac{1}{k}\pmatrix{\alpha & \beta & \gamma \cr \epsilon  & 1 & \exp
-i\lambda_1\cr \delta & \exp -i\lambda_2 & \exp -i\lambda_3 } 
\label{e2001}
\end{eqnarray}
By unitarity the values of the different elements are related to the phases
$\lambda_i$ (it is neccesary to suppose $\beta\not = 0$):
\begin{eqnarray}
\mid\beta\mid^2=\mid\gamma\mid^2=\mid\epsilon\mid^2=\mid\delta\mid^2&=&2\mid
\cos \frac{\Lambda}{2}\mid\nonumber \\
\kappa=\mid k\mid^2&=&2+\mid\beta\mid^2=4\sin^2\frac{\Lambda}{4}\nonumber \\
\mid \alpha\mid^2&=&2-\mid\beta\mid^2=4\cos^2\frac{\Lambda}{4}
\nonumber \\
\arg(\gamma-\beta)=\frac{\Lambda_+}{2};& &
\arg(\delta-\epsilon)=\frac{-\Lambda_-}{2}\nonumber\\
\cos(\Delta_\delta)&=&\pm\cos\frac{\Lambda}{4}=\pm\frac{\mid\alpha\mid}{2}\nonumber\\
\Lambda=\lambda_1+\lambda_2-\lambda_3;\quad
\Lambda_+&=&\lambda_1-\lambda_2+\lambda_3;\quad
\Lambda_-=\lambda_1-\lambda_2-\lambda_3\nonumber\\
\Delta_\delta &=& \arg(\beta+\epsilon-\alpha)\nonumber\\
\mid a_{CP}\mid
 \equiv 2\frac{\mid\Im(u_{11}u_{22}u_{12}^\ast u_{21}^\ast)\mid}{\mid
u_{11} u_{22}\mid^2+\mid u_{12} u_{21}\mid^2}
 &=& \frac{1}{4}\mid\alpha\mid \mid k\mid \nonumber
\end{eqnarray}

 The convention independent parameter
$a_{CP}$ \cite{jar1} gives a measure
of the intrinsic CP violation. Note that as long as two of the three masses are
degenerate or nearly degenerate the effective CP violation is in any case null
or very small.

{\bf For the case $\beta=0$}, the
mixing matrix has the trivial form
\begin{eqnarray}
u&=&\frac{1}{\surd 2}\pmatrix{\surd 2 \exp i\delta & 0 & 0 \cr 0 & 1 &  1\cr 0 & -1 & 1 } 
\label{e2002}
\end{eqnarray}
There is a decoupling between the first and the other generations. The mixing
between the second and the third one is rigorously  maximal.
This case, which it is not supported apparently by experimental evidence 
will appear later when we consider $e-\mu$ maximal mixing.

{\bf The transition probabilities} between weak states can be written as 
\beq
P_{\alpha\beta}=\mid tr \exp-iH_0 t\ W_{(\alpha\beta)}^\dagger\mid^2
\eeq   
where $H_0$ is taken diagonal and 
the set of matrices $W_{(\alpha\beta)}$ is defined by
\beq
W_{ij,(\alpha\beta)}=u_{\alpha i}^\ast u_{\beta j}
\eeq
The probabilities
$P_{\alpha\beta}$ are linear combinations of 
$c_{ij},s_{ij}=\cos,\sin(2.57 \Delta m_{ij}^2 L/E)$,
$\Delta m^2_{ij}=m_i^2-m_j^2$ and the
neutrino masses $m_i$ (eV), $L$ (m), $E$ (MeV).

From the matrix given by (\ref{e2001}), we compute the survival probabilities:
\begin{eqnarray}
P_{ee}&=&P_{ee}^\infty-2\frac{\kappa^2-6\kappa+8}{\kappa^2}(c_{12}+c_{13})+2\frac{\kappa^2-4\kappa+4}{\kappa^2}
c_{23}\\
P_{\tau\tau}=P_{\mu\mu}
&=&P_{\mu\mu}^\infty+2\frac{\kappa-2}{\kappa^2}(c_{12}+c_{13})+\frac{2}{\kappa}
c_{23} 
\end{eqnarray}
with
\begin{eqnarray}
P_{ee}^\infty&=& \frac{1}{\mid k\mid^4}(\mid
\alpha\mid^4+\mid\beta\mid^4+\mid\gamma\mid^4)=3-\frac{16}{\kappa}+\frac{24}{\kappa^2}\label{e3001}\\
P_{\tau\tau}^\infty=P_{\mu\mu}^\infty&=& \frac{1}{\mid
k\mid^4}(2+\mid\beta\mid^4)=1-\frac{4}{\kappa}+\frac{6}{\kappa^2} \label{e3002}
\end{eqnarray}

With the same mixing matrix, some of the transition probabilities are:
\begin{eqnarray}
P_{\overline{\mu}\ \overline{e},\mu e}&=&\frac{\kappa-2}{\kappa^2}\left
(\mid\alpha\mid^2+2+2\mid\alpha\mid\left (\cos (E_{12}\pm\Delta_{\delta})+\cos
(E_{13}\pm\Delta_\delta\pm\Lambda/2)\right )+2\cos(E_{23}+\Lambda/2)\right )\nonumber \\
P_{\mu \tau}&=&\frac{1}{\kappa^2}\left
(\kappa+2\mid \beta\mid^2\left
 (\cos (E_{12}-\Delta/2)+\cos
(E_{13}+\Delta/2)\right )+2\cos(E_{23}+\Delta)\right ) 
\end{eqnarray}

In the second formula the $+(-)$ sign corresponds to $\nu_\mu\nu_e
(\overline{\nu_\mu}\overline{\nu_e})$ transitions. $E_{ij}=\Delta m_{ij}^2$.
We note that all the $P_{\alpha\beta}$ depend only on one parameter ($\kappa$
or $\Lambda$).

For the interpretation of the data of any concrete experiment, it is neccesary
to average the $c_{ij},s_{ij}$'s taking into consideration the finite volume of
the detector; the space, time and energy resolutions; the energy dependent cross
sections and experimental detection efficiencies. Unfortunely the different
experimental groups publish little or unmanageable information about these
details (see \cite{ush1} for an exception, where the averaging weight function is
given in an explicit and practical way). 

The effect of the averaging can be approximated conveniently considering an
uniform distribution of the ratio $L/E$ in the range $[0,2L/E]$. This is the
prescription adopted in \cite{per1}. We have examinated many other
prescriptions, in particular eliminating the  $L/E\to 0$ region. We have checked
that the following results are independent of the averaging procedure which is
adopted.

\begin{figure}
\centering\hspace{2cm}
\epsfig{file=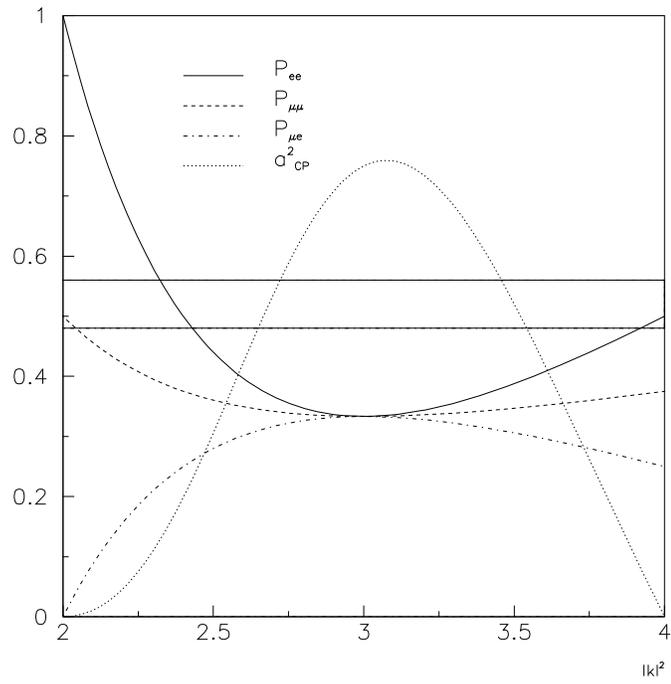,height=10cm}
\caption{The probabilities $P_{ee}^\infty,P_{\mu\mu}^\infty$ (eqs.
(\protect\ref{e3001}-\protect\ref{e3002})) and $\mid a_{CP}\mid^2$ as a function of
$\kappa=\mid k\mid^2$. The value $\mid k\mid^2=3$ correspond to the threefold
maximal mixing, here $P_{ee}^\infty=P_{\mu\mu}^\infty$.
The horizontal band correspond to the average of the solar
neutrino experiments: $P_{ee}^\infty=0.52\pm 0.04$.}
\label{f2}
\end{figure}

In the region of very high $L/E$, as in the solar neutrino 
case for not so small
$\Delta m^2$'s, the averaged $c_{ij},s_{ij}$ are always practically zero. The
transition probabilities become independent of $L/E$ there.
The statistical average of the values from the four solar neutrino experiments
(taken from Table (1) in \cite{per1}) is $P_{ee}=0.52\pm 0.04$.
In Fig.(\ref{f2}), we plot $P_{ee}^\infty$ 
as a function of $\kappa=\mid k\mid^2$.
we see that there is only a very restricted region in the 
$\kappa$ line allowed by the solar neutrino data.
If we request $P_{ee}\simeq 1/2$, then
there are two possible solutions $\kappa=2.4$ or $\kappa=4$.
If $P_{ee}\simeq 0.4-0.5$
then $\kappa\simeq 2.4-2.5$ or $\kappa\simeq 3.8-4$. If $P_{ee}$ is clearly
greater than 1/2 but less than $\approx 0.6$ then only one region is allowed: $\kappa\simeq2.3-2.4$.
The threefold maximal mixing correspond to the value $\kappa=3$. This case is
discarded by the solar data except if a particular strong mass hierarchy is
considered as done in \cite{per1}.

For definiteness, we take $\kappa=2.5$, in this case
\begin{eqnarray}
 \Lambda=3.6;\; \Lambda_\delta=2.2;\; & &\mid\alpha\mid^2=1.5;\;
\mid\beta\mid^2=0.5 
\nonumber
\end{eqnarray}

The mixing matrix is 
\begin{eqnarray}
u&=&\pmatrix{0.77 e^{i\delta} & 0.44 & 0.44 e^{-i\lambda/2}
 \cr
0.44 & 1 & 1 \cr 0.44 e^{i\lambda / 2}& e^{-i\lambda} & 1}
\label{e4001}\\ \nonumber\\
\lambda&=&3.65\ rad,\;   \delta=2.23\ rad  \nonumber
\end{eqnarray}

{\bf The allowed high $\kappa$ region} is only possible if
$P_{ee}^{\infty}\simeq 0.4-0.5$. In the limit $\kappa\to 4$, we have
$\mid\alpha\mid=0$. 
 The mixing matrix takes the form
($\lambda_3=0$  for convenience):
\begin{eqnarray}
u&=&\frac{1}{2}\pmatrix{0 & w & -sw \cr w & 1 & s 1\cr -s w & s 1 &  1 } 
\label{e1001}\\
w&=&\sqrt{2} \exp -i\delta \quad s=\pm 1
\nonumber
\end{eqnarray}
There are not CP violation,  $a_{CP}=0$.

We give explicitly the form of some of the matrices $W_{(\alpha\beta)}$ in this case:
\beq
\begin{array}{cc}
W_{(11)}=\frac{1}{2}\pmatrix{0&0&0\cr 0&1& -1 \cr 0 & -1 & 1}& \\[10mm]
  W_{(22)}=
\frac{1}{4}\pmatrix{2&w^\ast&s w^\ast\cr w&1& 
s 1 \cr s w & s 1 & 1}&
  W_{(33)}=
\frac{1}{4}\pmatrix{2&-w^\ast&-s w^\ast\cr -w&1& 
s 1 \cr  -sw & s 1 & 1}
\end{array}
\label{e4002}
\eeq

The probabilities become:
\begin{eqnarray}
P_{ee}&=&\frac{1}{2}(1+c_{23})\\
P_{\mu\mu}=P_{\tau\tau}
&=&\frac{1}{8}(3+2 c_{21}+2c_{31}+c_{32})\\
P_{\mu e}=P_{\overline{\mu}\overline{e}}
&=&\frac{1}{4}(1-c_{23})\\
P_{\mu\tau}&=&\frac{1}{8}(3-2 c_{21}-2c_{31}+c_{32})
\end{eqnarray}

{\bf The maximal mixing in the $e-\mu$ or $e-\tau$ sectors}. 
These cases are obtained trivially from the previous $\mu-\tau$ case.
$P_{ee}$ becomes
equal to the $P_{\mu\mu}$ which was obtained  before. According to Figure 
(\ref{f2}) (now to the curve labeled as $P_{\mu\mu}$) 
this case is incompatible with a large value for $P_{ee}$ ($>\approx
0.50 $).
If we force a value for $P_{ee}\approx 0.4$ (and subsequently
 we favour the Homestake data) then $\kappa\approx 2$ and
$\mid \beta\mid^2\approx 0$, we recover aproximately 
the matrix given by Eq.(\ref{e2002}).

{\bf Including matter effects}. It is important to show that matter effects does
not affect the solar probabilities, as they play an important role in our 
derivation. We can easily show that they effectively doesn't affect for the case
given by the mixing matrix (\ref{e1001}).
It was shown the same in  \cite{per2} for the fully maximal case. We expect for
the general case the
effect, if any, to be small also.
The amplitude matrix in an arbitrary basis can be written with generality as 
\beq
{\cal A}= \exp -i H_0 t\ {\cal A}_r 
\eeq
where the matrix $A_r$ accounts for any matter effect.
 The  transition probabilities between weak states can be written as 
\beq
P_{\alpha\beta}=\mid tr \exp-iH_0 t\ A_r\ W_{(\alpha\beta)}^\dagger\mid^2
\eeq   
with $W_{(\alpha\beta)}$  the same matrices
as before.

For the solar case $A_r$ can be further decomposed (see \cite{emi1,emi2}) as 
\begin{eqnarray}
A_{ij,r}=\delta_{ij}+W_{ij,(11)} A_{ij,s}
\end{eqnarray}
For the matrix (\ref{e4001}), the matrix
 $W_{(11)}$ is given by Eq.(\ref{e4002}).
So $A_r$ has the form
\begin{eqnarray}
A_r=\left [ \begin{array}{c|c} 1& 0\\ \hline 0 & A_{r}^{(2)} \end{array} \right ] 
\end{eqnarray}
and
\begin{eqnarray}
A_r\times W_{(11)}= 
\left [ \begin{array}{c|c} 0& 0\\ \hline 0 & A_{r}^{(2)}\times W_{(11)}^{(2)} \end{array} \right ] 
\end{eqnarray}
where the superindex indicates the restriction to a dim 2 matrix.
\begin{eqnarray}
P_{ee}=\mid tr e^{-iH_0 t} A_{r}^{(2)}\times W_{(11)}^{(2)}\mid^2
\end{eqnarray}

We have reduced the problem to a problem with two generations and maximal mixing
between them. As it can be seen directly from Eq.(36) 
in \cite{emi1}, the
probability in this case is independent of any value of the other parameters.

{\bf In Fig.(\ref{f1})} (top, lines) we plot $P_{ee},P_{\mu\mu}$ for different
cases (
for any case $P_{\mu\mu}$ is the lowest line),
always with the mass spectrum given by \cite{yas1} (except case D). We see there are not strong
differences among the different cases. 
The case D, where a different mass spectrum has been used, is  clearly not
satisfactory in the medium range.
Accordingly, in Fig.(\ref{f1}) (bottom), we plot transition probabilities. The
model
predicts the general trend of the data. The LSND point would situated itself in
the
transition region given by the mass scale $\Delta m_0^2$. However the low error
quoted in the last published result
makes very difficult to reconcile  model and experiment.

{\bf Conclusions and consequences for future experiments.}
We have seen that quasi-maximal mixing in the $\mu-\tau$ sector represents
a simple, highly predictive ansatz to describe the general trend
of all the experimental evidence  with  three generation neutrino oscillations.
Small local deviations, for example among the solar data, could be accounted
using additional  mechanisms: residual matter oscillation effects, magnetic
field effects. The introduction of a non-unitary, still quasi-maximal, mixing
matrix
should deserve further attention in order, for example, 
to understand better the LSND value.

 For the CHORUS and NOMAD accelerator experiments \cite{jong1,ast1} the value of the quotient
$L/E\approx 10^{-2}$. According to the scenario made plausible by this work they
fall in a region where very little signal ($P_{\mu\tau}\approx 0$) is expected.
The results (soon coming) of these experiments would allow to improve the
existing exclusion plots but they would have  a very little discovering
potential. That situation could improve if they were able to increase
significantly
their
efficiency for the lower energy neutrinos.

During the period 1996-1997 two new solar experiments are scheduled to start,
 Super-Kamioka and SNO (\cite{braewan}). Both experiments will be able to
measure the flux of solar neutrinos coming from the 
${}^8 B$ decay as a function of the neutrino energy. Any departure from the
theoretically well known flux could be a signal for neutrino oscillations. As
presented in this work, $P_{ee}$ is 
constant at large L/E. For a fixed energy, the average over $L$ alone would be
also enough to warranty the constancy of $P_{ee}$, at least for appropriate long
term measurements, and the neutrino energy spectrum could appear unmodified even
in presence of neutrino oscillations. However, as it is shown in \cite{bil1},
still the neutrino oscillations could have checkable visible consequences.
  
Long baseline experiments as Fermilab/Soudan or 
CERN/Gran Sasso with $L/E\approx 10^2-10^4$ offer the most
realistic expectation of being able to find positive evidence of
  neutrino oscillations. In the same sense, the upgrading of LSND increasing L
in a factor $\approx 10$ 
would be highly recomendable. Taking into account the Fig.(\ref{f1}) they would
enter in a region with with a discovering potential nearly guaranteed.
The rapid increase of the transition probability here could make inneccesary
even
further improvements in the collimation of its neutrino beam. 

\vspace{1cm}

{\Large{\bf Acknowledgments.}}

I would like to thank to Peter Minkowski for many enlighthening discussions. This 
work has been supported in part by the Wolferman-Nageli Foundation (Switzerland)
and by the MEC-CYCIT (Spain).

\newpage

\begin{table}
\centering
\begin{tabular}{|c|c|c|c|}
\hline
Experiment  & L/E (m/MeV)& P(90\%) & Ref. \\
\hline
KARMEN &  17.7 m/40 MeV $=0.4$ & $P_{\overline{\nu_\mu}\ \overline{\nu_e}}$ $<3.1\
10^{-3}$  &\protect\cite{arm1}\\
BNL &  1 km/1-4 GeV $=0.2-1$ & 
$P_{\mu e}$,$P_{\overline{\mu}\ \overline{e}}$ $<1.5\
10^{-3}$  & \protect\cite{bor1}\\
LSND &  30 m/30-60 MeV $=0.5-1$ & $P_{\overline{\mu}\ \overline{e}}$ $=3.1\pm
1.0\pm 0.5\ 10^{-3}$  
& \protect\cite{atha1}\\
E531 &  $\approx 0.01-0.04$ & $P_{\nu_\mu\nu_\tau}$ $<2
\ 10^{-3}$  &\protect\cite{ush1}\\
CHORUS/NOMAD &  0.8 km/30 GeV $=0.03$  & $P_{\mu\tau}$  & \protect\cite{jong1,ast1}
\\ \hline
\end{tabular}
\caption{Results for the transition probabilities measured in diverse reactor
experiments.(CHORUS and NOMAD have not presented results so far).}
\label{t1}
\end{table}

\clearpage

\newpage

%\bibliography{a}

\begin{thebibliography}{10}



\bibitem{per1}
{ P.~Harrison, D.~Perkins, and W.~Scott}, 
RAL-94-125.

\bibitem{bil1}
{ S.~Bilenky, C.~Giunti, and C.~Kim},
 hep-ph/9602383.
\bibitem{per2}
{ P.~Harrison, D.~Perkins, and W.~Scott}, 
 RAL-TR-95-078.


\bibitem{koi1}
{ Y.~Koide},
  hep-ph/9603376.
\bibitem{fuku1}
{ Y.~Fukuda et~al.},
 Phys. Lett. B., 335 (1994), pp.~237--245.

\bibitem{atha1}
{ C.~Athanassopoulos et~al.},
 Phys. Rev. Lett., 75
  (1995), pp.~2650--2658. For the latest update: nucl-ex/9605001.
\bibitem{yas1}
{ O.~Yasuda}, 
hep-ph/9602342.


\bibitem{pdg1} PDG. {\em Rev. Part. Properties}.
Phys. Rev. 50(1994)3, pp.~1700--1710.

\bibitem{jar1}
{ C.~Jarlskog},  Phys. Rev. Lett., 55 (1985), pp.~1039--1045.
\bibitem{ush1}
{ N.~Ushida et~al.}, 
Phys. Rev. Lett., 57 (1986),
  pp.~2897--2899.

\bibitem{emi1}
{ E.~Torrente-Lujan}, 
Phys. Rev. D., 53 (1996), pp.~53--67.

\bibitem{emi2}
{ E.~Torrente-Lujan}, 
hep-ph/9602398.




\bibitem{arm1}
{ B.~Armbruster et~al.},  Nucl. Phys. (Proc. Suppl.) B, 38 (1995),
  pp.~235--240.

\bibitem{bor1}
{ L.~Borodosky et~al.}, 
Phys. Rev. Lett., 68 (1992), pp.~274--298.

\bibitem{jong1}
{ M.~de~Jong et~al.},  CERN-SPSLC/91-131,90-42.


\bibitem{ast1}
{ P.~Astier et~al.}, CERN-SPSLC/91-21,91-48,P261.

\bibitem{braewan}
G.T. Ewan, {\em Proc.  $4^{th}$ Int. W. on Neut.
Telescopes}, Venezia, March 1992, ed. by M. Baldo Ceolin.


\end{thebibliography}
%\bibliographystyle{siam}

\end{document}